\documentclass[epj,nopacs]{svjour}
\usepackage{graphics}
\usepackage{epsfig}
\usepackage{amssymb}
\begin{document}
\title{Hard collisions of photons: plea for a common language}
\author{Ji\v{r}\'{\i} Ch\'{y}la
%
}
\institute{Institute of Physics
of the Academy of Sciences of the Czech Republic\\
Na Slovance 2, Prague 8, Czech Republic}
%
\date{Received: date / Revised version: date}
%
\abstract{
An attempt is made to sort out ambiguities existing in the current usage
of several basic concepts describing hard collisions of photons. It is
argued that appropriate terminology is often a prerequisite for correct
physics.
%
} 
\maketitle
\section{Motivation}
\label{sec:intro}
Hard collisions of quasireal as well as virtual photons have recently
received large experimental and theoretical attention, mainly due to the
fact they offer new ways of testing perturbative QCD. The novel feature
of such tests arises from the fact that the photon exhibits two
apparently different faces: it acts as structureless
particle and simultaneously as hadron-like object. At very short
distances photon looks simpler than hadrons, but at experimentally
accessible ones its hadron-like properties are essential.

This novel and intriguing aspect of photon physics has, however, also
lead to misunderstanding and confusion resulting mostly from
unsettled or inappropriate terminology. It is perhaps not a coincidence
that just those aspects which distinguish hard collisions of photons
from those of hadrons involve ambiguous, unsuitable or obsolete notions
and notation. Different names are used to denote the same content, but
also conversely, a particular term is employed by different people to
express different contents. The main purpose of this paper is to
contribute to defining a set of notions and definitions which is
unambiguous, sufficient but not redundant, and as exact as possible.

\section{Basic facts}
\label{sec:basic}
The two-facet appearance of the photon is due to the existence of the
point-like coupling of photons to quark-antiquark pairs, described
by pure QED. This coupling generates the inhomogeneous terms on the
r.h.s. of the evolution equations
\begin{eqnarray}
\frac{{\mathrm d}\Sigma(x,M)}{{\mathrm d}\ln M^2}& =&
\delta_{\Sigma}k_q+P_{qq}\otimes \Sigma+ P_{qG}\otimes G,
\label{Sigmaevolution}
\\ \frac{{\mathrm d}G(x,M)}{{\mathrm d}\ln M^2} & =& k_G+
P_{Gq}\otimes \Sigma+ P_{GG}\otimes G, \label{Gevolution} \\
\frac{{\mathrm d}q_{\mathrm {NS}}(x,M)}{{\mathrm d}\ln M^2}& =&
\delta_{\mathrm {NS}} k_q+P_{\mathrm {NS}}\otimes q_{\mathrm{NS}},
\label{NSevolution}
\end{eqnarray}
where
$\delta_{\mathrm{NS}} \equiv 6n_f\left(\langle e^4\rangle-\langle
e^2\rangle ^2\right)$, $\delta_{\Sigma}=6n_f\langle e^2\rangle$ and
\begin{eqnarray}
\Sigma(x,M) & \equiv &
\sum_{i=1}^{n_f} \left[q_i(x,M)+\overline{q}_i(x,M)\right],
\label{singlet}\nonumber\\
q_{\mathrm{NS}}(x,M) & \equiv &
\sum_{i=1}^{n_f}\left(e_i^2-\langle e^2\rangle\right)
\left(q_i(x,M)+\overline{q}_i(x,M)\right),
\label{nonsinglet}\nonumber
\end{eqnarray}
describing the dependence of parton distribution functions (PDF)
on the {\em factorization scale} $M$. To order $\alpha$ the splitting
functions $P_{ij}$ and $k_i$ are given in powers of $\alpha_s(M)$
\begin{eqnarray}
k_q(x,M) & = & \frac{\alpha}{2\pi}\left[k^{(0)}_q(x)+
\frac{\alpha_s(M)}{2\pi}k_q^{(1)}(x)+\cdots\right],
\label{splitquark} \nonumber\\ k_G(x,M) & = &
\frac{\alpha}{2\pi}\left[
\frac{\alpha_s(M)}{2\pi}k_G^{(1)}(x)+
\frac{\alpha_s^2(M)}{2\pi}k^{(2)}_G(x)+\cdots\;\right],
\label{splitgluon}\nonumber \\ P_{ij}(x,M) & = &
\frac{\alpha_s(M)}{2\pi}P^{(0)}_{ij}(x) +
\frac{\alpha_s^2(M)}{2\pi} P_{ij}^{(1)}(x)+\cdots,
\label{splitpij}\nonumber
\end{eqnarray}
where $k_q^{(0)}(x)=x^2+(1-x)^2$ and $P^{(0)}_{ij}(x)$ are
unique while all higher order splitting functions
$k^{(j)}_q,k^{(j)}_G,P^{(j)}_{kl},j\ge 1$ depend on the choice of
the {\em factorization scheme} (FS). The equations
(\ref{Sigmaevolution}-\ref{NSevolution}) can alternatively be
rewritten as evolution
equations for $q_i(x,M),\overline{q}_i(x,M)$ and $G(x,M)$.

The structure function $F_2^{\gamma}(x,Q^2)$ is given as
\begin{eqnarray}
\frac{1}{x}F_2^{\gamma}(x,Q^2)= q_{\mathrm{NS}}(M)\otimes
C_q(Q/M)+\frac{\alpha}{2\pi}\delta_{\mathrm{NS}}C_{\gamma}+ & &
\nonumber \\ \langle e^2\rangle \Sigma(M)\otimes
C_q(Q/M)+\frac{\alpha}{2\pi} \langle
e^2\rangle\delta_{\Sigma}C_{\gamma}+& &\nonumber\\
 \langle e^2\rangle G(M)\otimes C_G(Q/M),~~~~~~~~~~~~~~~~~~~~~
  & &
\label{S+Gpart}
\end{eqnarray}
where the coefficient functions $C_q,C_G,C_{\gamma}$ can be expanded in
powers of $\alpha_s$ taken at the {\em renormalization scale}
$\mu$:
\begin{eqnarray}
C_q(x,Q/M) & = & \delta(1-x)~~~~+
~~~\frac{\alpha_s}{2\pi}C^{(1)}_q(x, Q/M)+\cdots,
\label{cq}\nonumber \\
C_G(x,Q/M) & = & ~~~~~~~~~~~~~~~~~~~~~
\frac{\alpha_s}{2\pi}C^{(1)}_G(x,Q/M)+\cdots,
\label{cG} \nonumber\\
C_{\gamma}(x,Q/M) & = &
C_{\gamma}^{(0)}(x,Q/M)+
\frac{\alpha_s }{2\pi}C_{\gamma}^{(1)}(x,Q/M)+\cdots,
\label{cg}\nonumber
\end{eqnarray}
where $\kappa(x)\equiv 8x(1-x)-1$ and $\alpha_s\equiv \alpha_s(\mu)$.
The lowest order contribution to $C_{\gamma}$
\begin{equation}
C_{\gamma}^{(0)}(x,Q/M)=(x^2+(1-x)^2)\ln\frac{Q^2(1-x)}{M^2x}+
\kappa(x)
\label{c0}
\end{equation}
comes, similarly as $k_q^{(0)}(x)$, from pure QED, which provides
the lowest order contribution to $F_2^{\gamma}$ in the form
\begin{eqnarray}
\frac{1}{x}F_2^{\gamma,\mathrm{QED}}(x,Q^2)&=
&\sum_{i=1}^{n_f}e_i^2\left(q_i^{\mathrm{QED}}(x,Q)+
\overline{q}_i^{\mathrm{QED}}(x,Q)\right)+\label{q}\nonumber\\
& &\frac{\alpha}{2\pi}6n_f\langle e^4\rangle C_{\gamma}^{(0)}(x,1),
\label{lowest}
\end{eqnarray}
where the quark distribution functions $q_i^{\mathrm{QED}}(x,M)$ satisfy
the evolution equations (\ref{Sigmaevolution}-\ref{NSevolution}) with
the purely QED inhomogeneous splitting function $k_q^{(0)}$ only. The
above formulae hold for $n_f$ massless quark flavors, while for
heavy quarks quark mass effects have to be taken into account.

\section{Basic notions}
\label{sec:vocabulary}
In this Section I will review some of the notions and notation used
for the description of hard scattering of photons, discussing their
various connotations and overlaps.

\subsection{Direct \& resolved photon}
\label{subsec:dirres}
Careful attention deserves already the interpretation of the basic
concepts ``direct'' and ``resolved'' photon. For instance, in
\cite{sasgamma} one finds the following expression for the photon
``wave function''
\begin{eqnarray}
\lefteqn{\mid\!\gamma\rangle=
c_{\mathrm{bare}}\mid\!\gamma_{\mathrm{bare}}\rangle +}& &\label{state}\\
& & \sum_Vc_V\mid\!V\rangle+\sum_qc_q
\mid\!{\mathrm{q}\overline{\mathrm{q}}}\rangle
+\sum_lc_l\mid\!l^+l^-\rangle,\nonumber
\end{eqnarray}
where the first sum runs over vector mesons, the second over
$q\overline{q}$ pairs and third over analogous pairs of leptons and
antileptons. The coefficients
$c_l^2\approx (\alpha/2\pi)\ln(\mu^2/m_l^2)$ and
$c_q^2\approx (\alpha/2\pi)\ln(\mu^2/k_0^2)$ depend on
the scale $\mu$ ``used to probe the photon'' and satisfy
the ``unitarity'' relation
\begin{equation}
c_{\mathrm{bare}}^2=1-\sum_{V} c_V^2-\sum_q c_q^2-\sum_l c_l^2.
\label{unitarity}
\end{equation}
In the language of \cite{sasgamma} the first term on the r.h.s. of
(\ref{state}) defines the direct photon, whereas the remaining
ones correspond to the resolved (either to partons or leptons) photon.
The separation (\ref{state}) of the photon state looks intuitively
appealing, but must not be taken literally. In fact both
relations (\ref{state}) and (\ref{unitarity}) make sense only as
shorthand for the statements concerning {\em cross sections} of the
processes involving the initial photon. The terms ``direct'' and
``resolved'' are in fact not adjectives of the {\em state}
of the photon, but of its {\em interactions}.

\begin{figure}\centering
\epsfig{file=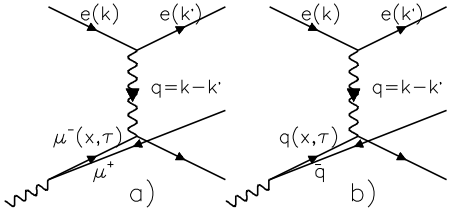,width=6cm}
\caption{Feynman diagrams describing in pure QED the contributions of
$\mu^+\mu^-$ and $q\overline{q}$ pairs to the cross section of DIS of
electrons on the photon. Momentum fraction and virtuality of quarks
and leptons entering the elastic scattering with an electron are
denoted as $x$ and $\tau$.}
\label{qedfig}
\end{figure}

To illustrate this point in detail, let us consider the perturbatively
calculable pure QED contribution to $F_2^{\gamma}(x,Q^2)$, i.e. we
discard in (\ref{state}) and (\ref{unitarity}) the contributions of
vector mesons and take into account only electromagnetic interactions
of quarks. The exact cross section for the lepton-antilepton
production in DIS of electrons on the photon, described by the diagram
in Fig. \ref{qedfig}a, is given to order $\alpha$ and assuming
$Q^2\gg m_l^2$ as
\begin{equation}
\frac{\mathrm{d}\sigma(e^-\gamma\rightarrow e^-l\overline{l})}
{{\mathrm{d}}x{\mathrm{d}}Q^2}=
\frac{2\pi\alpha^2}{xQ^4}F_{2,l}^{\gamma}(x,Q^2)\left(1+(1-y)^2\right),
 \label{dsigma}
\end{equation}
where
\begin{equation}
F_{2,l}^{\gamma}(x,Q^2)=\frac{\alpha}{2\pi}2e_l^4x\left[k_q^{(0)}(x)
\ln\frac{Q^2(1-x)}{m_l^2x}+\kappa(x)\right]
\label{F}
\end{equation}
and with the replacement $e_l^2\rightarrow 3e_q^2$ similarly for the
$q\overline{q}$ pair production. In QED eq. (\ref{F}) gives the exact
result, but even there it makes sense to separate it into the parts
\begin{eqnarray}
F_{2,l}^{\gamma,{\mathrm{res}}}(x,Q^2)& \equiv&
\frac{\alpha}{2\pi}2e_l^4x\left[k_q^{(0)}(x)
\ln\frac{M^2}{m_l^2}\right],\label{flres}\\
F_{2,q}^{\gamma,{\mathrm{res}}}(x,Q^2)& \equiv&
\frac{\alpha}{2\pi}6e_q^4x\left[k_q^{(0)}(x)
\ln\frac{M^2}{m_q^2}\right],\label{fqres}
\end{eqnarray}
coming from region of almost colinear
$\gamma\rightarrow l\overline{l}$
and $\gamma\rightarrow q\overline{q}$ splitting and the rest
\begin{eqnarray}
\!\!\!\!\!\!\!
F_{2,l}^{\gamma,{\mathrm{dir}}}(x,Q^2)&\equiv&
\frac{\alpha}{2\pi}2e_l^4x\left[k_q^{(0)}
\ln\frac{Q^2(1-x)}{M^2x}+\kappa(x)\right]\label{fldir}\\
\!\!\!\!\!\!\!F_{2,q}^{\gamma,{\mathrm{dir}}}(x,Q^2)&\equiv&
\frac{\alpha}{2\pi}6e_q^4x\left[k_q^{(0)}
\ln\frac{Q^2(1-x)}{M^2x}+\kappa(x)\right].\label{fqdir}
\end{eqnarray}
The ``large log'' $\ln(M^2/m_{l}^2)$ in (\ref{flres}) results from
integrating the singular part, proportional to $1/\tau$, of the cross
sections $\mathrm{d}\sigma(e^-\gamma\rightarrow e^-l\overline{l})/
\mathrm{d}x\mathrm{d}Q^2\mathrm{d}\tau$ over the region of small
lepton virtuality $\tau$ (see Fig. \ref{qedfig}), the factorization
scale $M^2$ defining the upper limit of this integration. The
remaining part of this integral, depending on both $M^2$ and $Q^2$,
together with the integral over the whole phase space of the regular
part, yields (\ref{fldir}). Analogously for the $q\overline{q}$
production described by (\ref{fqres}) and (\ref{fqdir}), the latter
generating $C_{\gamma}^{(0)}$ in (\ref{c0}).

Defining generic lepton and quark distribution functions of the photon
as
\begin{eqnarray}
l^{\mathrm{QED}}(x,M)& \equiv & \frac{\alpha}{2\pi}e_l^2
k_q^{(0)}(x)\ln\frac{M^2}{m_l^2}, \label{leptonssinQED}\\
q^{\mathrm{QED}}(x,M)& \equiv & \frac{\alpha}{2\pi}3e_q^2
k_q^{(0)}(x)\ln\frac{M^2}{m_q^2} \label{quarksinQED}
\end{eqnarray}
allows us to express $F_2^{\gamma,\mathrm{res}}$ in terms of
quark and lepton distribution functions in the same way as $F_2$ for
hadrons.
The coefficients $c_l^2,c_q^2$ appearing in (\ref{state}) are thus in
fact just lepton and quark distributions functions $l(x,M)$ and $q(x,M)$
defined above, except that in the latter case $m_q$ is replaced by the
phenomenological parameter $k_0$ and the QCD correction (discussed in
the next subsection) are included. Expressing $c_{\mathrm{bare}}^2$ as
in (\ref{unitarity}) is thus nothing but a shorthand for defining the
direct photon contribution, i.e. the sum of (\ref{fldir}) and
(\ref{fqdir}), by subtracting the resolved photon contributions
(\ref{flres}-\ref{fqres}) from the full expression for $F_2^{\gamma}$
\begin{equation}
F_{2}^{\gamma,{\mathrm{dir}}}=
F_{2}^{\gamma}-F_{2,l}^{\gamma,{\mathrm{res}}}-
F_{2,q}^{\gamma,{\mathrm{res}}}
\label{dirdef}
\end{equation}
and similarly for other physical quantities.

Although the quark and lepton distribution functions
do not characterize the state of the photon, they are {\em universal},
i.e. do not depend on the hard process in which the virtual quarks or
leptons in Fig. \ref{qedfig} are involved. This crucial property implies
that quark and lepton distribution functions are {\em attributes} of the
photon and provides the basis for the predictive power of these concepts.
In QED the decomposition of $F_{2}^{\gamma}$ into the resolved
(\ref{flres}-\ref{fqres})) and direct (\ref{fldir}-\ref{fqdir}) photon
contributions is actually not necessary since the full result (\ref{F})
is known. Nevertheless, such decomposition useful even there, because
it shows clearly the origin of the concept of quark and lepton
distribution functions and illustrates the central fact that they
describe cross sections rather then states!

Switching on QCD implies several modifications of the QED expression
(\ref{lowest}):
\begin{itemize}
\item adding perturbative QCD corrections to quark
distribution functions and introducing the gluon distribution function,
\item adding perturbative QCD corrections to direct photon contribution
$F_{2,q}^{\gamma,{\mathrm{dir}}}$, i.e. generating
$C_{\gamma}^{(i)},i\ge 1$,
\item adding QCD corrections to quark and gluon coefficient
functions $C_q$ and $C_G$,
\item including the so called hadron-like contribution, discussed in
below,
\end{itemize}
but the physical meaning of PDF of the photon remains basically the
same as in pure QED. Recall that for hadrons the very notion of PDF is
based on the {\em factorization theorem}, which is a statement about
cross sections. This theorem relies in turn on the validity of the KLN
theorem, which guarantees absence of mass
singularities in the sums of cross sections over the sets of degenerate
initial and final states. The formal mathematical analogy between
the UV renormalization of QCD and ``IR renormalization group'' technique
used in \cite{CFP} to define PDF of hadrons must not disguise the fact
that the former deals with basic quantities of QCD lagrangian (fields,
masses and charges), whereas the latter with cross sections of physical
processes. Moreover, the standard UV renormalization of QCD actually
{\em precedes} the factorization procedure. In other words, all fields,
masses and charges entering the factorization procedure are
{\em renormalized} quantities!

\subsection{Factorization theorem and the ``bare'' PDF}
\label{bare}
In \cite{CFP} the ``bare'' PDF of the photon are introduced but in this
case the adjective ``bare'' has nothing to do with the
standard UV renormalization of fields, masses and charges, and concerns
IR behaviour of PDF, i.e. cross sections. Specifically, the unknown,
nonperturbative ``bare'' PDF of hadrons are {\em assumed} to contain
mass singularities which, according to KLN theorem, exactly cancel those
due to homogeneous perturbative splitting of incoming and outgoing
partons
\footnote{For discussion of this important point see \cite{Field}.}.
Mass singularities of the ``bare'' PDF appear when we evaluate, as
prescribed by the KLN theorem, cross sections of multiparton initial
states, like for instance, an electron scattering on a {\em pair} of
partons with parallel momenta $p_1,p_2$, degenerate with a single
parton with momentum $p_1+p_2$. The absorption of mass singularities of
cross sections coming from perturbative splitting of single incoming
partons in the ``dressed'' PDF is just an equivalent, and simpler, way
of describing the result of such procedure. Nevertheless, for bound
states the validity of such cancellation is nontrivial.

For the photon the same mechanism can be expected to operate
for the ``hadron-like'' part of PDF, but not for the point-like one
\footnote{See the next Subsection for definition of these notions.}.
For the latter the parallel logs resulting from the
primary $\gamma\rightarrow q\overline{q}$ splitting are not cancelled
by the singularity of the ``bare'' PDF, but cut-off
by the confinement at $M_0$, analogously as in QED, where, however,
this cut-off is provided by quark masses. The inherent ambiguity in the
choice of $M_0$ then naturally relates the point-like and hadron-like
parts of full PDF. Identifying the direct photon contributions,
with the ``bare'' photon, as done, for instance in \cite{stefan}, is
thus flawed.

The different nature of the UV renormalization of QCD quantities and
IR ``renormalization'' of PDF is also the main argument for keeping the
factorization and renormalization scales $M$ and $\mu$ as independent
free parameters. The former sets the {\em upper} bound on
the virtualities of quantum fluctuations taken into account, via the
factorization theorem, in the definition of PDF, whereas the latter
determines the lower bound on virtualities included in the renormalized
charges, masses and fields. There is no reason, why these
two scales should be identified.

\subsection{Point-like \& hadron-like}
\label{subsec:point}
The terms {\em point-like} (PL) and {\em hadron-like} (HAD) have
been used by the GRV group \cite{grv1,grv2,grv3,grs1,grs2}
to describe the separation  of a general solution of the
evolution equations (\ref{Sigmaevolution}-\ref{NSevolution})
into the particular solution of the full
inhomogeneous equations and a general solution of the corresponding
homogeneous one. A subset of the former resulting from the resummation
of series of diagrams in Fig. \ref{figpl}, which start with the
purely QED vertex $\gamma\rightarrow q\overline{q}$ and vanish at
$M=M_0$, are called {\em point-like} solutions. Due to the fact
that $M_0$ is in principle
\footnote{Though not in practice if we want to describe the data.
For instance, $M_0=0.6$ GeV in SaS1D parameterizations, whereas
$M_0=2$ GeV in SaS2D ones},
arbitrary parameter, the separation of
quark and gluon distribution functions into their point-like and
hadron-like parts is, however, ambiguous. In general
we can thus write ($D=q,\overline{q},G$)
\begin{equation}
D(x,M)= D^{\mathrm {PL}}(x,M,M_0)+D^{\mathrm{HAD}}(x,M,M_0).
\label{separation}
\end{equation}
The main difference between these two components
concerns their virtuality dependence. Whereas the hadron-like parts
fall-off with $P^2$ rapidly and essentially independently of $M^2$,
like $(M_0^2/P^2)^2$, the point-like ones decrease much more slowly
like $\ln(M^2/P^2)$. Quantitative aspects of the separation
(\ref{separation}) are discussed in \cite{prd}.

\subsection{QED \& QPM}
\label{qpm}
\begin{figure}\centering
\epsfig{file=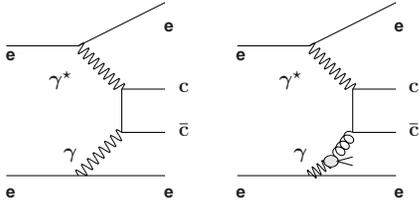,width=6cm}
\caption{Examples of leading order diagrams contributing to direct (left)
and single resolved photon parts of $F_{2,c\overline{c}}^{\gamma}$.}
\label{ccbar}
\end{figure}

Another notion often used in photon physics is the ``Quark-Parton
Model'' (QPM) contribution. For $\gamma\gamma$ processes
it usually stands for the lowest order, purely QED contribution
involving neither the PDF of the photon nor $\alpha_s$. For instance,
for heavy quark production in $\gamma\gamma$ collisions it comes from
the diagram in the left part of Fig. \ref{ccbar}, taken from
\cite{richard}, and similar diagram describes the lowest order, purely
QED contributions to jet production
in $\gamma\gamma$ collisions as well. In both cases the denomination
``QED'' contribution is certainly accurate and unambiguous.
The term QPM might seem appropriate for those processes,
for which the lowest order contributions do involve PDF of the photon,
but the lowest order parton level matrix
element are independent of $\alpha_s$, like, for instance, Drell-Yan
dilepton production in double resolved photon contribution. But as
nowadays all PDF used in such calculations do incorporate QCD effects
in their scale dependence, the term ``lowest order'' QCD contribution
is certainly more appropriate than ``QPM''.
Parton model had played indispensable role in the formulation of
QCD, but is now so firmly embedded therein that there is little reason
to use the denomination ``QPM'' for historical reasons only, when more
appropriate terms are available.

\subsection{Resolved photon \`{a} la DELPHI}
\label{subsec:DELPHI}
\begin{figure}\centering
\epsfig{file=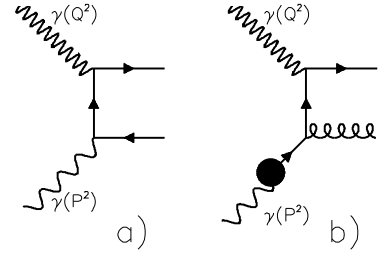,width=5cm}
\caption{Diagrams describing QPM (a) and single resolved photon (b)
contributions to $F_2^{\gamma}(x,Q^2)$ according to the terminology
used by the DELPHI Collaboration.}
\label{delphi}
\end{figure}

Unfortunately, there is no universal agreement on the content of even
the very basic notion of resolved photon contribution. Take, for
instance, the photon structure function $F_2^{\gamma}(x,Q^2)$ as
measured at LEP. While OPAL, L3 and ALEPH use this term in the sense
introduced in the Section \ref{subsec:point}, DELPHI \cite{Teddy,Sarka}
associates
it with the diagram in Fig. \ref{delphi}b, which corresponds to the
convolution of PDF of the target photon with the cross section
$\sigma_{\gamma q}$ of the process $\gamma(Q^2)+q\rightarrow q+G$,
cut-off at $p_T^{\mathrm{min}}\simeq 2$ GeV. In the standard terminology
the regular part of
$\mathrm{d}\sigma_{\gamma q}/\mathrm{d}\tau$ integrated
over the whole phase
space of the emitted gluon plus the integral over the singular part from
$M^2$ up to $Q^2$ gives the term proportional to $q\otimes C_q^{(1)}$ in
(\ref{S+Gpart}), whereas the integral over the singular part up to $M^2$
is included in $q(x,M)$ and contributes to its scale dependence. On the
other hand, and again contrary to the standard procedure, the ``single
resolved'' photon contribution of Fig. \ref{delphi}b is summed with what
DELPHI calls ``QPM'' contribution of Fig. \ref{delphi}a, regularized
by means of (constituent) quark masses $m_u=m_d=0.3$ GeV, $m_s=0.5$
GeV, $m_c=1.5$ GeV and $m_b=4.5$ GeV. Note that in standard
terminology the singular part of this contribution is included in
the point-like part of quark distribution functions of the photon and
becomes thus part of the resolved photon contribution, whereas its
regular part goes to $C_{\gamma}^{(0)}$ and describes the lowest order
direct photon contribution.

The fact that the DELPHI defines the ``resolved photon'' contribution
to $F_2^{\gamma}$ by this non-standard way is unfortunate, but the real
problem with their treatment of $\gamma\gamma$ collisions is the way
they simulate genuine hadron-like (called ``VMD'' by DELPHI)
contribution to $F_2^{\gamma}(x,Q^2)$. Note that as their sum of QPM
and single resolved photon contributions contains both the basic QED
contribution and some QCD corrections to it, it is at least related
to the conventional single resolved photon contribution. For the genuine
hadron-like part of PDF of the photon the diagram in Fig. \ref{delphi}b
cut-ff at $p_t^{\mathrm{min}}\simeq 2$ GeV describes ${\cal O}(\alpha_s)$
correction to basic DIS process e$+$q$\rightarrow$e$+$q, but not
this lowest order contribution itself! Consequently, MC event generator
TWOGAM used by DELPHI differs substantially from, for instance, HERWIG
or PYTHIA event generators at $x\lesssim 0.01$, where the genuine
hadron-like components of the photon PDF dominate.

\subsection{Point-like \& hadron-like: alternative use}
\label{subsec:alternative}
The terms {\em hadron-like} and {\em point-like} are used by
some theorists \cite{laenen,laenen2} as well as experimentalists
\cite{richard,stefan} in a different sense than as introduced above,
namely in the sense of {\em resolved} and {\em direct} photon
contributions. This usage relies on formal mathematical
similarity between the expressions for cross sections of hard collisions
of hadrons and the resolved photon. The fact that PDF of the photon
satisfy different evolution equations than those of hadrons is from this
point of view of secondary importance.

The choice of terminology is a matter of convention, but we
should avoid the present situation, where the terms ``hadron-like'' and
``point-like'' are used in two different senses. As there are good
reasons for separating PDF of the photon into their genuine hadron-like
and point-like parts, despite the inherent ambiguity of such separation,
some notions should exist for this purpose. Since there is little
justification to denote as ``hadron-like'' the contributions that have
manifestly nothing to do with the existence and properties of hadrons,
the terminology used in \cite{grv1,grv2,grv3,grs1,grs2} is in my view
preferable to that of \cite{richard,laenen,laenen2,stefan}. Moreover,
as we shall see in Section \ref{subsec:example}, the absence of unique
interpretation of the term ``hadron-like'' may lead to unnecessary
weakening of important experimental observations.

\subsection{Anomalous, VMD, bare: who needs them?}
\label{subsec:bare}
The terminology introduced so far exhausts all concepts necessary for
the description of hard collisions of photons, but three other
terms, {\em VMD, anomalous and bare} photon are also widely used.
However, for one reason or another, these notions are poor alternatives
to the terms {\em hadron-like, point-like} and {\em direct}.

The term {\em anomalous part} of PDF of the photon is particularly
unsuitable substitute for the term {\em point-like part} introduced
above. The term {\em anomalous} itself
has been coined in \cite{kingsley,zerwas2} to denote the result of a
simple (from current point of view) calculation first done in
\cite{zerwas} of purely QED contribution to cross section of the
process $\gamma^*(Q^2)\gamma(0)\rightarrow{\mathrm{hadrons}}$ based
on the box diagram. The fact that the resulting contribution
to $F_2^{\gamma}(x,Q^2)$, coinciding with (\ref{F}) for $\kappa=0$,
turned out to
be proportional to $\ln Q^2$ was rightly considered ``essentially
different'' \cite{zerwas} from experimental results on deep inelastic
scattering on hadrons, which showed approximate $Q^2$-independence.
This latter
observation, combined with the idea of Vector Meson Dominance had
naturally lead to the expectation of a similar scaling behaviour for
$F_2^{\gamma}(x,Q^2)$. Recall that \cite{zerwas} was written a few month
before the birth of QCD, at the time when parton model was still in
its infancy. The term ``anomalous'' was thus introduced to denote the
behavior anomalous with respect to exact scaling of parton model
predictions for proton structure function $F_2^{\mathrm{p}}(x,Q^2)$.
In QCD the logarithmic scaling violations are, on the other hand,
commonplace and, moreover, stem from the same origin as those found
in \cite{zerwas} for $F_2^{\gamma,\mathrm{QED}}$. From current point
of view the term ``anomalous'' describes nothing anomalous,
but on the contrary the behavior of $F_2^{\gamma}$ which results
from the standard QED coupling of photons to pairs of quarks and
antiquarks. Despite its historical connotation, I see no compelling
reason for retaining the term {\em anomalous} when the more
appropriate term {\em point-like} is available and, indeed, used by
part of physics community.

The hadron-like parts of PDF of the photon are often claimed (see, for
instance, \cite{grv1,grs1}) to be modeled by PDF of vector mesons and
therefore called ``VMD''. However, as there is no experimental
information on PDF of vector mesons, the latter are actually
approximated by those of pions, extracted from analyses of Drell-Yan
processes in $\pi$p collisions. In view of huge differences between the
role of vector mesons and pions in the Standard model, demonstrated
among other things by large difference between the masses of pions and
vector mesons, this further assumption is, however, difficult to
justify. In any case what is assumed for this part of PDF is a rather
general shape expected for meson states and the term {\em hadron-like}
is thus clearly more appropriate.

As for the term ``bare'', I have argued already in Section
\ref{subsec:dirres} why it has no role in the description of hard
collisions of the photon and can only cause confusion when used as
substitute for ``direct''.

\section{Graphical representation}
\label{sec:graphics}
Proper graphical representation of hard processes involving photon in
the initial state is complicated by the interplay between the point-like
part of the resolved photon contribution and the direct photon one.
Before going into details, let me emphasize that the factorization
theorem, on which the very concept of PDF is based, concerns cross
sections, whereas conventional Feynman diagrams describe individual
contributions to the corresponding amplitudes.
\begin{figure}\centering
\epsfig{file=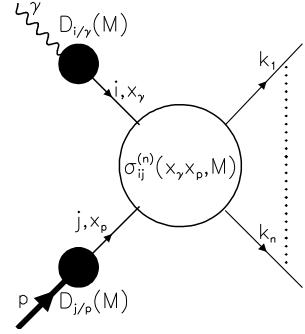,width=4.5cm}
\caption{Graphical representation of the factorization theorem for
hard processes in $\gamma$p collisions. The lines representing beam
remnants are not drawn.}
\label{factor}
\end{figure}
For instance, Fig. \ref{factor} represents graphically the expression for
the resolved
photon contribution to inclusive cross section for the production
of $n$-parton final state in $\gamma$p collisions, which has a generic form
\begin{equation}
\sigma^{(n)}_{\gamma{\mathrm{p}}}=\sum_{i,j}D_{i/\gamma}(M)\otimes
\sigma_{ij}^{(n)}(M)\otimes D_{j/{\mathrm{p}}}(M),
\label{sigmagp}
\end{equation}
where parton level cross sections are given as
\begin{equation}
\sigma_{ij}^{(n)}(M)=
\alpha_s^{\kappa}\left(\sigma_{ij}^{(n)}({\mathrm{LO}})+
\alpha_s\sigma^{(n)}_{ij}({\mathrm{NLO}})+\cdots\right)
\label{sigmas}
\end{equation}
and $\kappa\ge 0$.
The problem is that except for $\sigma_{ij}^{(n)}({\mathrm{LO}})$,
all higher order cross sections in (\ref{sigmas}) involve
integrals over the unobserved partons with subsequent
subtraction of singular terms, as well as addition of loop corrections,
which have different number of final state partons than the tree
diagrams. As a result, only the lowest order cross section
$\sigma_{ij}^{(n)}({\mathrm{LO}})$ can meaningfully be attached to the
blobs representing PDF of the beam particles, as in Fig. \ref{sasfig}a.

Whereas for hadrons the problem with proper graphical representation
appears first at NLO, for photon induced hard processes it appears once
the resolved photon contribution is taken into account.
The conventional way of graphical representation of
quark distribution functions of the photon (see, for instance, Fig. 1a,c
of \cite{sas}), reproduced in Fig. \ref{sasfig}a,b, combines the
solid blob, representing their hadron-like part and the standard Feynman
diagram vertex $\gamma\rightarrow q\overline{q}$, standing for the
point-like one. However, this representation of the point-like part of
the resolved photon contributions is unsatisfactory because the
diagram in Fig. \ref{sasfig}b plays in fact double role.
The evaluation of its contribution to the cross section for dijet
production involves integration over the virtuality $\tau$, which is
split into two parts in the manner described in Section
\ref{subsec:dirres}.
Consequently, part of the contribution of this diagram goes into
the definition of the point-like part of quark distribution function and
is thus included in the resolved photon contribution, whereas the other
one defines the NLO direct photon one. The diagram in Fig.
\ref{sasfig}b cannot therefore be meaningfully associated to either the
direct or resolved photon contribution.
\begin{figure}\centering
\epsfig{file=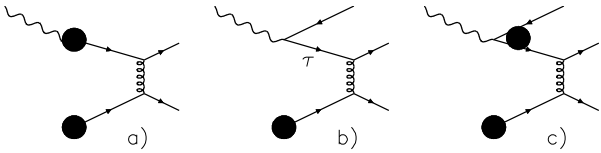,width=8cm}
\caption{Conventional representation of the lowest
order resolved photon contribution corresponding the hadron-like (a) and
point-like (b) parts of quark distribution function of the photon
involved in hard collision with a quark from any beam particle
(lower blob). Graphical representation of the point-like part suggested
in \cite{FS} is shown in (c).}
\label{sasfig}
\end{figure}
The representation of the point-like part of quark distribution functions
depicted in Fig. \ref{sasfig}b disregards also the fact that the
point-like parts of quark distribution functions include resummation of
the effects of multiple parton emissions off the primary $q\overline{q}$
pair. Finally, the point-like part exists also for the gluon distribution
function of the photon, whereas Fig. \ref{sasfig}b concerns quarks only.

\begin{figure}\unitlength 1mm
\begin{picture}(100,33)
\put(0,22){\epsfig{file=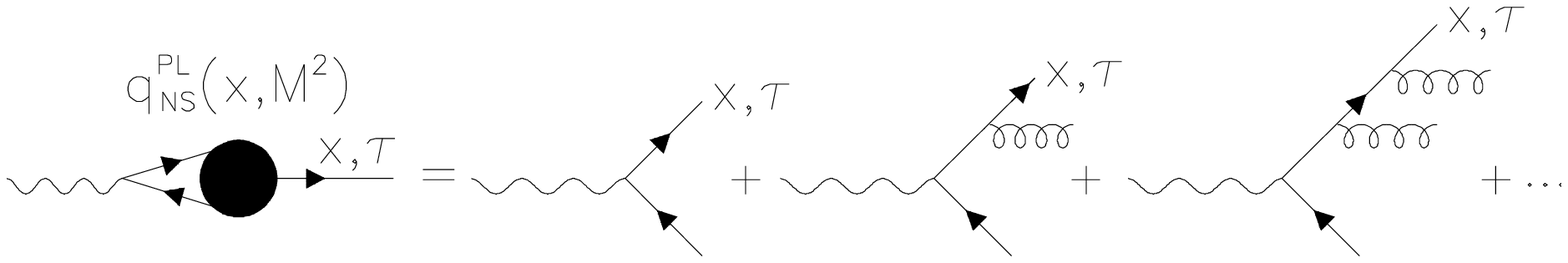,width=8cm}}
\put(0,5){\epsfig{file=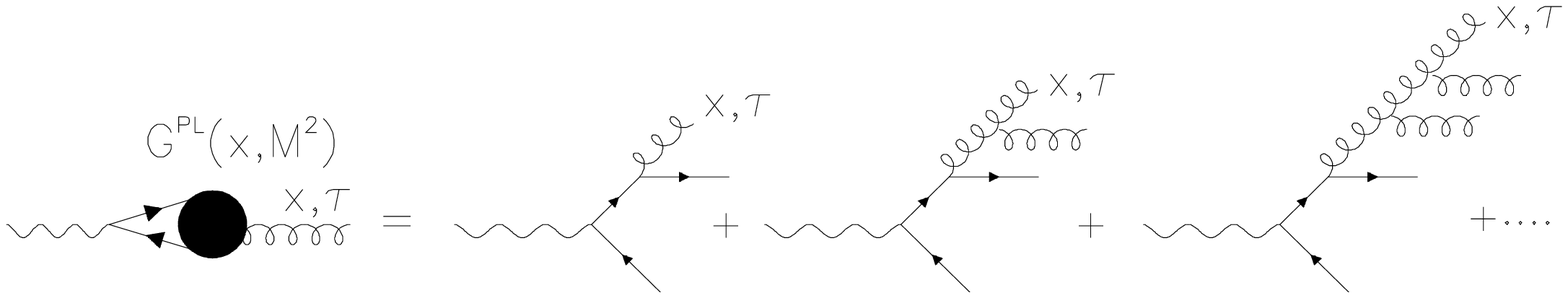,width=8cm}}
\end{picture}
\caption{Diagrams defining the point-like parts of nonsinglet quark and
gluon distribution functions. The resummation involves integration over
parton virtualities $\tau\le M^2$ and is represented by the junction
of the blob and the $\gamma\rightarrow q\overline{q}$ vertex. Partons
going into the hard collision are denoted by $x,\tau$.}
\label{figpl}
\end{figure}

Recently, however, the authors of \cite{FS} have come up with a good
idea (see Fig. \ref{sasfig}c) how to represent graphically the point-like
part of quark distribution functions of the photon which reflects its
primary QED origin .
Albeit basically correct, this suggestion goes only half-way in solving
the problem of appropriate graphical representation of the point-like
parts of PDF of the photon as it concerns point-like parts of quark
distribution functions only. The obvious extension of this idea
is to represent the point-like parts of PDF of the photon resulting
from the resummation of cross sections corresponding to diagrams in
Fig. \ref{figpl} by the special blobs on the l.h.s. of the equation
signs in this figure for both quarks and gluons.
Moreover, I suggest discarding the lines representing beam particle
remnants as the blobs themselves can take up their role. The full solid
blobs would be reserved for genuine hadron-like parts of PDF of the
photon.
\begin{figure}\centering
\epsfig{file=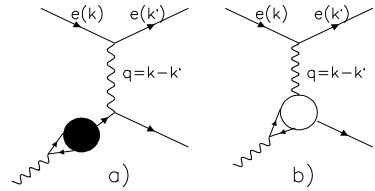,width=5cm}
\caption{Proposed representation of the point-like part of the resolved
photon (a) and direct (b) contributions to $F_2^{\gamma}$.}
\label{myfig}
\end{figure}
As, however, the point-like parts of quark distribution functions of
the photon always appear accompanied by the corresponding direct photon
contribution of the same order, we have to invent appropriate graphical
representation of the latter as well. Because the direct photon
contributions are process dependent, any such representation must
connect the incoming photon not only to outgoing parton but also to the
hard collision itself. For $F_2^{\gamma,{\mathrm{dir}}}$ possible such
representation is suggested in Fig. \ref{myfig}b, along with the diagram
a) describing the contribution of the point-like part of the resolved
photon. The open blob connecting the initial photon and final quark to
the exchanged probing photon represents graphically this process
dependence and simultaneously suggests its relation to the solid blob
in Fig. \ref{myfig}a. Let me reiterate that the above graphical
representations are meaningful for the LO resolved photon contribution
only.

There are, however, processes, like heavy quark production in
$\gamma\gamma$ collisions, which do not involve at the lowest order of
$\alpha_s$ the resolved photon contribution. In such cases the direct
photon contribution is described by a simple Feynman diagram, like in
the left part of Fig. \ref{ccbar}.

\section{``Leading'' and ``next--to--leading'' orders}
\label{sec:lonlo}
Existing QCD analyses of hard collisions of photons are burdened by
the lack of clear separation of genuine QCD effects from those
of pure QED origin. Take, for example, heavy quark production in
$\gamma\gamma$ collisions discussed in \cite{qqbar}. Counting, as in
\cite{laenen,laenen2}, the lowest order, purely QED contribution of
the left diagram in Fig. \ref{ccbar} as the ``LO QCD'' contribution
is legitimate, but implies that the content of the term ``NLO QCD
approximation'' is different for $F^{\gamma}_{2,c\overline{c}}$ than,
for instance, the analogous $F_2^{\mathrm{p}}$. To perform QCD analysis
of $F_2^{\mathrm{p}}$ in a well-defined renormalization scheme requires
working within the NLO (or higher) QCD approximation. Using the
terminology of \cite{laenen,laenen2} the ``NNLO QCD''analysis of
$F^{\gamma}_{2,c\overline{c}}$ would be required for the analysis of
$F^{\gamma}_{2,c\overline{c}}$ in a well-defined renormalization scheme!

To avoid misunderstanding it is in my view preferable to discard for
the purpose of defining the terms ``leading-order'' and
``next-to-leading order'' QCD analysis the pure QED contributions.

\section{Photon remnant}
\label{sec:remnant}
The {\em photon remnant} is another concept requiring careful use
because it is employed in two distinct meanings.
First, it simply denotes ``something'' flying roughly in the
direction of the incoming photon. Used in this sense, {\em photon
remnant} has similar content as {\em proton remnant}, except that
the mean transverse momentum of partons making up the photon
remnant is bigger for the point-like part than the hadron-like
part of PDF of the photon. However, this difference is to large
extent washed out by hadronization effects and there is thus
little noticeable difference between the properties of photon
remnant in these two classes of events.
\begin{figure}\centering
\epsfig{file=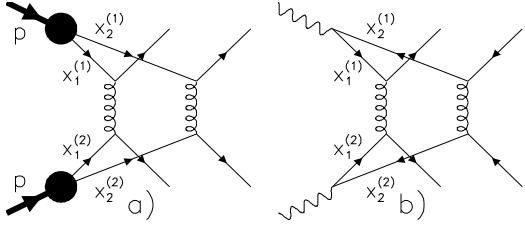,width=7cm}
\caption{Double parton scattering in proton-proton collision (a)
and photon-proton collision involving the leading contribution
to the point-like part of PDF of the photon (b).}
\label{remfig}
\end{figure}

{\em Photon remnant} plays a specific role within the framework of
multiparton interactions (MI) models. Proper treatment of this
additional source of soft particles requires distinguishing the
hadron-like and point-like parts of PDF of the photon since for the
additional partonic collisions beam remnants play the role of incoming
particles. The assumption of {\em uncorrelated multiple scatters},
which lies at the heart of the MI model \cite{Tor,Mike}, amounts to
assuming the factorization of the general multiparton distribution
functions into products of single parton ones. This might be a good
approximation for hadrons, or hadron-like component of the photon, at
small to moderate $x$, but is less justified for the point-like one.
Indeed, in both analyses \cite{Tor,Mike} and their Monte-Carlo
implementations in HERWIG and PYTHIA, multiple scatters are simulated
only in events corresponding to hadron-like parts of PDF of the photon.

However, multiple parton scatters make in principle sense even for the
point-like part of PDF of the photon.
The difference between the MI model for hadrons and pointlike part of the
photon is indicated in Fig. \ref{remfig}, which
shows Feynman diagrams describing double parton interaction
in pp and $\gamma\gamma$ collisions,
in the latter case coming from the point-like parts of PDF of both
photons. In pp collisions the standard way of simulating double parton
scattering is to pick up from each of the protons any possible
pair of partons $p^{(1)}_k,p^{(2)}_k,k=1,2$ with momentum fractions
$x_k^{(1)},x_k^{(2)},k=1,2$. For instance, both partons from each of
the protons in Fig. \ref{remfig}a can be quarks with the same small
momentum fractions $x_1^{(j)}\simeq x_2^{(j)}, j=1,2$. In hadron-hadron
collisions the two pairs of partons involved in double parton
scattering are thus uncorrelated as far as their identity as well as
momentum fractions $(x_1^{(1)},x_1^{(2)}),(x_2^{(1)},x_2^{(2)})$ are
concerned (apart form the condition $x_1^{(k)}+x_2^{(k)}\le 1,k=1,2$).

For contributions of the point-like parts of PDF of the photon the
above approximation is manifestly invalid for the dominant part of the
quark distribution functions, coming from the primary QED splitting
$\gamma\rightarrow q\overline{q}$. For this contribution, depicted in
Fig. \ref{remfig}b,  both the parton species and momentum fractions are
fully correlated as only $q\overline{q}$ pairs from both photons are
allowed and, moreover, $x_2^{(k)}=1-x_1^{(k)},k=1,2$. For the full
point-like part of PDF of the photon this correlation is somewhat
washed out, but it is clear that double parton distribution functions
do not factorize into the product of single ones as assumed in standard
formulation of the MI model.

\section{Fluctuating photon}
\label{sec:fluctuation}
The concept of ``fluctuating photon'' is well-defined only
within the framework of dispersion relations written first in
\cite{bjorken} for moments of $F_2^{\gamma}(x,P^2,Q^2)$.
In \cite{sas1} virtuality dependence of PDF of the photon was introduced
using generalization of these dispersion relations to virtuality
dependence of PDF $f_a(x,P^2,M^2)$ themselves
\begin{eqnarray}
\lefteqn{f_a^{\gamma}
(x,M^2,P^2)=}& & \nonumber\\
& & \sum_{V}\left(\frac{m_V^2}{P^2+m_V^2}\right)^2
\frac{4\pi\alpha}{f_V^2}f_a^{\gamma,V}(x,M^2,M_0^2)\label{dispersion}\\& &
+\int_{M_0^2}^{M^2}\!\frac{\mathrm{d}k^2}{k^2}\!\left(\frac{k^2}{k^2+P^2}
\right)^2\!\frac{\alpha}{2\pi}\!\sum_q2e_q^2
f_a^{\gamma,\mathrm{q}\overline{\mathrm{q}}}(x,M^2,k^2),
\nonumber
\end{eqnarray}
where the first sum runs over the vector
mesons and the function
$f_a^{\gamma,\mathrm{q}\overline{\mathrm{q}}}(x,M^2,k^2)$
satisfies standard homogeneous evolution equation with the boundary
condition
\begin{displaymath}
f_a^{\gamma,\mathrm{q}\overline{\mathrm{q}}}(x,k^2,k^2)=
3\left(x^2+(1-x)^2\right)(\delta_{aq}+\delta_{a\overline{q}}).
\end{displaymath}
Let me emphasize at this point that the difference between the VMD
(hadron-like in my terminology) and anomalous (point-like) is not,
as claimed for instance in \cite{gg}, in the sign of the off-shellness
of the $q\overline{q}$ pair to which the initial space-like photon
couples. In both components the off-shellness of the $q\overline{q}$
pair is exactly the same as that of the original initial photon, i.e.
negative.

The initial space-like photon involved in hard collisions at LEP and
HERA interacts by first coupling to $q\overline{q}$ pairs of the same
virtuality $P^2$ as the photon itself. The dispersion relations
(\ref{dispersion}) allow us to
express PDF (or, in general, cross sections) of these states as the
sum of contributions associated with two types of singularities in the
time-like region of the target photon virtuality $P^2$: discrete set of
poles corresponding to vector mesons and continuum cut corresponding to
free $q\overline{q}$ pairs. The phrase ``photon fluctuates'' make sense
only as a shorthand for the preceding statement. There is also
no principal difference between the contributions of vector mesons, in
\cite{dainton} associated with the ``valence-driven structure'', and
those of the free $q\overline{q}$ pairs, in \cite{dainton}
giving the ``fluctuation-driven'' contribution. The only difference is
that latter is perturbatively calculable, whereas the former is not.

It is also worth emphasizing that the association \cite{dainton} of the
``fluctuation driven'' (point-like in my terminology) contributions to
$F_2^{\gamma}$ with the $\ln Q^2$ pattern of scaling violations holds
for $\gamma_T^*$ only. The scale dependence of the point-like part of PDF
of $\gamma_L^*$ is typically hadron-like \cite{gammal}, i.e. without this
$\ln Q^2$ behaviour, yet it arises from exactly same ``fluctuation
driven'' mechanism as the $\ln Q^2$ behavior of point-like part of PDF
of $\gamma_T^*$.

\section{OPAL results on the structure of photon}
\label{subsec:example}
I will now discuss three recent OPAL papers bringing new information
on the structure of the photon. Although the results of all three papers
are phrased in similar words as evidence about the ``hadronic'' or
``hadron-like'' component of the photon, their true importance and
impact is substantially different.

\subsection{Jet production in $\gamma\gamma$ collisions}
\label{subsec:opaljets}
In \cite{opaljets} cross sections for dijet production in $\gamma\gamma$
collisions at $\sqrt{s_{\mathrm{ee}}}=161$ and $172$ GeV were measured
in the kinematical region $E_T\ge 3$ GeV and $x_{\gamma}\gtrsim 0.1$ and
confronted with the LO MC event generators PYTHIA and PHOJET
using GRV and SaS1D PDF of the photon. The data require the
presence of double resolved photon contribution, but do not cover
$x_{\gamma}\lesssim 0.01$ where the genuine hadron-like parts of PDF of
the photon are expected to dominate. The conclusion of \cite{opaljets},
namely that photons ``appear resolved through its fluctuations into
hadronic components'' therefore implies that the term ``hadronic'' is
used in the sense of \cite{laenen,laenen2}, i.e. as a substitute for
``resolved''.

\subsection{$F^{\gamma}_2(x,Q^2)$ at low $x$.}
\label{subsec:opallowx}
In \cite{opallowx} $F_2^{\gamma}(x,Q^2)$ has for the first time been
measured for moderate $1.5\le Q^2\le 30$ GeV$^2$ and very small
$x\gtrsim 0.002$.
\begin{figure}\centering
\epsfig{file=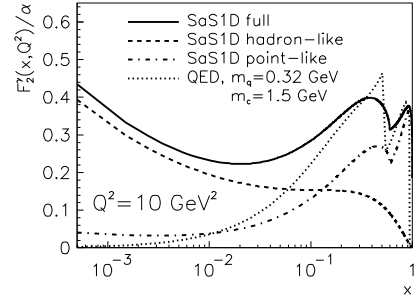,height=4cm}
\caption{Individual contributions to $F_2^{\gamma}$ at low $x$.}
\label{lowxfig}
\end{figure}
Comparison with existing parameterizations as well as pure QED
contribution show convincingly, that the data are far above the
contribution of the point-like parts of PDF of the photon and below
$x\simeq 0.05$ definitely require the presence of the genuine hadron-like
part, roughly the size given by GRV or SaS1D parameterizations! This is
illustrated in Fig. \ref{lowxfig}, where the SaS1D results are
compared with pure QED ones for $Q^2=10$ GeV$^2$ and taking into account
four quarks with masses $m_u=m_d=m_s=0.32$ GeV, $m_c=1.5$ GeV. Noting
that the OPAL data (not shown) are roughly in agreement with the solid
curve we conclude that although at very small $x_{\gamma}$ even the
point-like contribution starts to rise (the rise by itself is thus no
evidence for the genuine hadron-like part of PDF of the photon), its
{\em magnitude} insufficient to account for the data! The conclusion of
this paper: ``These results show that the photon must contain a significant
hadron-like component at low $x$'' thus concerns important evidence about
the genuine hadron-like part of the photon, and not, as that of
\cite{opaljets}, merely about the resolved photon contribution.

\subsection{Charm contribution to $F^{\gamma}_2(x,Q^2)$}
\begin{figure}\centering
\epsfig{file=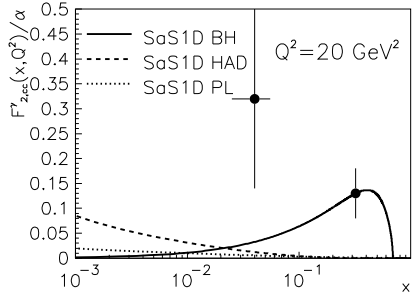,height=4cm}
\caption{Bethe-Heitler, genuine hadron-like and light quark initiated
point-like contributions to $F_{2,c\overline{c}}^{\gamma}$.}
\label{F2cfig}
\end{figure}

In Fig. \ref{F2cfig} the measurement \cite{richard} of
$F^{\gamma}_{2,c\overline{c}}$ at $\langle\!Q^2\!\rangle=20$ GeV$^2$ is
compared to pure QED calculation using exact Bethe-Heitler formula for
$m_c=1.5$ GeV, as well as LO single resolved photon contribution
evaluated with SaS1D parameterization of PDF of the photon and plotted
separately for their hadron-like and point-like parts.
In the region $x\simeq 0.05$ the data are significantly above all these
calculations, as well as those (not shown) including NLO QCD effects in
the single resolved photon contribution and ${\cal O}(\alpha_s)$
corrections to the Bethe-Heitler cross section in the direct photon
channel. Although the error bar is large, it seems that in this case
even including the genuine hadron-like contribution may not be
sufficient to describe the OPAL data.

The term {\em hadron-like} component is used in \cite{richard} again as
an equivalent of the {\em resolved} photon contribution. Thus, although
it is true that \cite{richard} ``the measurement suggests a non-zero
hadron-like component of $F^{\gamma}_{2,c\overline{c}}$.'', the data
tell us in fact much more. Unfortunately, the importance of this message
is somewhat diluted by the used terminology.

The above discussion of three recent OPAL papers underlines the need for
unambiguous terminology adopted, if not generally, than at least within
one Collaboration.

\section{Summary and conclusions}
\label{sec:summary}
Here is my proposal how to describe hard collisions of photons. Use the
terms
\begin{itemize}
\item {\em resolved} and {\em direct} to distinguish
 contributions that do and do not involve PDF of the photon,
\item {\em hadron-like} and {\em point-like} to
 distinguish two components of a general solution of the corresponding
 evolution equations, the latter coming from resummation of
 perturbative contributions of multiple parton emissions,
\item {\em QED} to denote the contributions that involve
 neither $\alpha_s$ nor PDF of the photon,
\item {\em LO and NLO} to denote approximations that contain first and
 first two nonzero powers of $\alpha_s$ in hard scattering cross
 sections or splitting functions.
\end{itemize}
Avoid, on the other hand, notions {\em bare, anomalous, VMD, QPM}, which
are, for one reason or another, less suitable for description of hard
collisions of the photon than those listed above.

\vspace*{1cm}
I am grateful to J. Form\'{a}nek, M. Krawczyk, S. Maxfield, R. Nisius,
B. P\"{o}tter, K. Sedl\'{a}k, T. Sj\"{o}strand, S. S\"{o}ldner-Rembold,
M. Ta\v{s}evsk\'{y}, \v{S}. Todorovov\'{a}-Nov\'{a}$~$ and T. Wengler for
comments and suggestions. This work was supported by the Ministry of
Education of the Czech Republic under the project LN00A006.

\end{document}